\documentclass[runningheads,a4paper]{llncs}

\usepackage[cmex10]{amsmath}
\usepackage{amssymb}
\usepackage{graphicx}
\usepackage{wrapfig}
\usepackage{subfig}
\usepackage{extdash} %
\usepackage{float}
\usepackage[svgnames]{xcolor}
\usepackage{marvosym}

\usepackage{url}
\urldef{\mailsa}\path|peleska@uni-bremen.de|
\urldef{\mailsb}\path|aeha@dtu.dk|
\urldef{\mailsc}\path|mario.gleirscher@uni-bremen.de|

\newcommand{\keywords}[1]{\par\addvspace\baselineskip
\noindent\keywordname\enspace\ignorespaces#1}

\usepackage{xspace}

\usepackage{hyperref}

\makeatletter
\begin{document}

\mainmatter  %

\title{Probabilistic Risk Assessment of an Obstacle Detection System for  GoA~4 Freight Trains}

\titlerunning{Risk Assessment of an Obstacle Detection System}

\author{
Mario Gleirscher\inst{1,3}\orcidID{0000-0002-9445-6863} \and
   Anne E.~Haxthausen\inst{2}\orcidID{0000-0001-7349-8872} \and 
   Jan Peleska\inst{1}\orcidID{0000-0003-3667-9775}\thanks{Partially funded by the German Ministry of Economics,
      Grant Agreement~20X1908E.} 
}

\authorrunning{Gleirscher, Haxthausen, and Peleska}

\institute{
University of Bremen,
Department of Mathematics and Computer Science,
Germany\\
{\tt \{mario.gleirscher,peleska\}@uni-bremen.de}
\and
DTU Compute, Technical University of Denmark, Kongens Lyngby, Denmark\\
\mailsb
\and
Autonomy Assurance International Programme, University of York,
Deramore Lane, York YO10-5GH, United Kingdom
}

\maketitle

\begin{abstract}
  In this paper, a quantitative risk assessment approach is discussed
  for the design of an obstacle detection function for low-speed
  freight trains with grade of automation (GoA)~4.  In this 5-step
  approach, starting with single detection channels and ending with a
  three-out-of-three (3oo3) model constructed of three independent
  dual-channel modules and a voter, a probabilistic assessment is
  exemplified, using a combination of statistical methods and
  parametric stochastic model checking. It is illustrated that, under
  certain not unreasonable assumptions, the resulting hazard rate
  becomes acceptable for specific application settings.
  The statistical approach for assessing the residual risk of
  misclassifications in convolutional neural networks and conventional
  image processing software suggests that high confidence can be
  placed into the safety-critical obstacle detection function, even
  though its implementation involves realistic machine learning
  uncertainties.
\end{abstract}

\section{Introduction}\label{sec:intro}

\subsubsection*{Motivation and Background}

Autonomous transportation systems, their technical feasibility, safety and security are currently in the main focus of both academic research and industrial developments. This has been caused by both the  significant progress made in academia   regarding the enabling technologies -- in particular, artificial intelligence (AI) --  and   the attractive business cases enabled by driverless road vehicles, trains, and aircrafts. 

A major obstacle preventing the immediate deployment of autonomous transportation systems in their designated operational environments is their safety assessment. The latter poses several technical challenges~\cite{DBLP:journals/itsm/KoopmanW17,RR-1478-RC,10.1145/3542945}, in particular, the trustworthiness of AI-based methods involving machine learning (such as obstacle recognition on roads and railway tracks), as well as standardisation challenges: in the railway and aviation domains, approved standards for the certification of safety-critical autonomous trains or aircrafts are still unavailable.  

The standardisation situation is more advanced in the automotive domain, where a stack of standards involving 
ISO~26262~\cite{iso26262-4},   ISO~21448~\cite{iso21448}, and   ANSI/UL~4600\xspace~\cite{UL4600} have been approved by the US-American Department of Transportation for the certification of autonomous road vehicles.\footnote{See \url{https://www.youtube.com/watch?v=xCIjxiVO48Q}.}

The standard ANSI/UL~4600\xspace is of particular interest, since its authors emphasise that it is applicable to operational safety assurance on system level for {\it all} types of autonomous products~\cite[Section~1.2.1]{UL4600}, 
potentially with a preceding system-specific 
revision of the checklists proposed in the standard. In a previous publication, 
Peleska et al.~\cite{isolapaper2022} have suggested  a particular  control system architecture for autonomous trains and performed   a qualitative evaluation according to 
ANSI/UL~4600\xspace. This work resulted in the assessment that a system-level certification based on  this standard is feasible for the class of autonomous metro trains and freight trains, running in an open  operational environment, as can be expected 
in European railway networks today. 

It should be noted that   autonomous trains in closed environments (platform screen doors, secured tracks, as provided by underground metro trains or airport  people movers,  where the problem of unexpected obstacles can be neglected) already exist since decades~\cite{flammini_rssrail_2022}. The current challenge consists in integrating autonomous train operation safely into  the ``normal'' open operational environments of today's European railway networks.

\subsubsection*{Objectives and   Contributions}

This paper complements our previous work \cite{isolapaper2022}
with respect to probabilistic risk assessment and associated verification methods for an autonomous train control system architecture with the highest grade of automation GoA~4 (no train engine driver or other personnel present): for a real-world certification, it is necessary to 
add a probabilistic risk analysis to the qualitative evaluation. We specialise here on autonomous low-speed freight trains travelling across railway networks, the latter providing movement authorities via interlocking systems and radio block centres. For this type of train, the most important safety-relevant AI-based sub-system is the  \emph{obstacle detection (OD) module}, consisting of sensors and perceptors. As explained in the previous work~\cite{isolapaper2022}, the reaction to obstacle indications from OD and the state transitions between fully autonomous GoA4 mode and degraded modes due to sensor and perceptor failures can be specified, designed, implemented, verified, validated, and certified  with conventional methods, typically according to standards like EN~50128, EN~50129~\cite{CENELEC50128,CENELEC50129}. The OD function can be evaluated according to   ANSI/UL~4600\xspace, as explained in~\cite{isolapaper2022}.  

Note that while OD has been extensively   researched in the automotive domain~\cite{CARO2023102872}, the results obtained there cannot be directly transferred to railways considered in this paper: the two domains require OD for different distances, and the detection criteria differ, because trains require the obstacle association with their tracks.
We consider the following methodological aspects and risk analysis results to be the main contributions in this paper.
\begin{enumerate}
\item We propose a new verification method for (deep) neural networks (NN) performing classification tasks such as obstacle detection. This method allows to determine the residual probability ${p_\text{\Lightning}}$
for a systematic classification error in the trained NN. Increasing the training effort, this method allows us to  reduce  ${p_\text{\Lightning}}$ to an acceptable value. 

\item We use a redundant design for the obstacle detection (OD) function introduced in our previous work~\cite{isolapaper2022} that allows to reduce the probability of a detection failure, due to stochastic independence between redundant channels. To show stochastic independence, a new statistical method is proposed.

\item We use parametric 
stochastic  model checking   to obtain probabilistic results for the hazard rate of the OD function.
The parametric approach allows us to leave some     values undefined, so that their influence on the hazard rate becomes visible, and the concrete risk values can be calculated at a later point, when reliable concrete parameter values are available.

\item The probabilistic analysis shows that, using a redundant 3oo3 design where each of the three sub-systems consists of a dual-channel module, the OD function is already certifiable today with an acceptable hazard rate of less than $10^{-7}/h$ for low-speed autonomous freight trains, even if only camera-based sensors and perceptors are used.\footnote{The   requirement for low speed (less or equal 120km/h) is based on the fact that no reliable failure probabilities for camera-based obstacle detection modules have been published for trains with higher velocities~\cite{ristic-durrant_review_2021}.} Further reduction of the hazard rate can be achieved by using additional fail-stop sensors/perceptors based on different technologies (such as radar and LiDAR), and apply sensor/perceptor fusion over the results of the non-failing units.
\end{enumerate}

To the best of our knowledge, our contribution is the first to apply this combination of statistical tests
and stochastic model checking to the field of risk analysis for concrete designs of autonomous train control systems.

\subsubsection*{Overview}

The redundant design for the OD sensor/perceptor function
proposed in our previous work~\cite{isolapaper2022}  is presented again in Section~\ref{sec:background}. In Section~\ref{sec:THR}, the risk modelling objectives and the applicable tolerable hazard rate are discussed.
In Section~\ref{sec:riskmodellingapproach}, the statistical test strategies and the concept of risk modelling and probabilistic analysis by means of parametric stochastic model checking are described.   The results of the probabilistic analysis are presented. 
Section~\ref{sec:conc} contains a conclusion.
Below, we give references to related work where appropriate.

\section{Fail-Safe Design of Obstacle Detection Modules}\label{sec:background}

As described in our previous work~\cite{isolapaper2022}, the OD function cannot be validated according to the existing 
EN~5012x standards, since the latter do not consider AI-based functions whose behaviour depends on machine learning techniques, such as (deep) neural networks (NN). Consequently, the safety of the intended functionality not only depends on the software implementation (of neural networks), but also on the choice of  training data and validation data sets~\cite{iso21448}. 

\begin{figure}[t]
\begin{center}
\includegraphics[width=.8\textwidth,angle=0,origin=c]{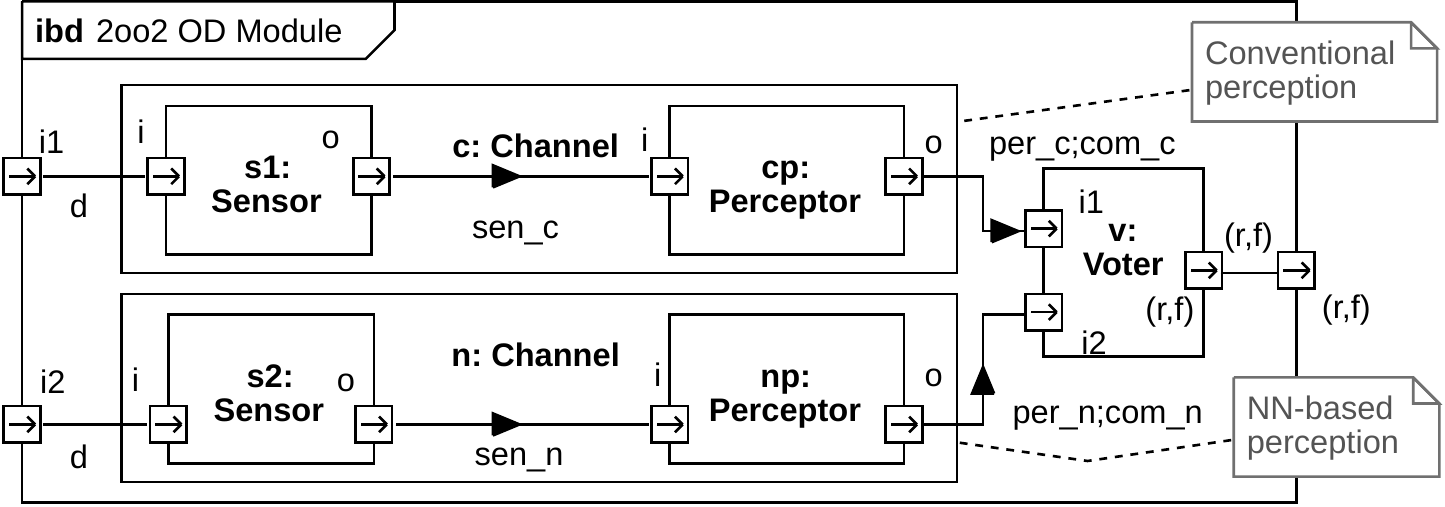}
\end{center}
\caption{2oo2 design pattern   for OD module or similar sense/perceive functions}
\label{fig:twochan}
\end{figure}
 
The standard ANSI/UL~4600\xspace provides guidance on
how the safety of the intended functionality should be demonstrated in a certification undertaking. From the performance data available (see discussion in Section~\ref{sec:tle-calc}), however, we   conclude that non-redundant sensor/perceptor components relying on machine learning and neural networks alone  cannot achieve the   tolerable hazard rates for
safety integrity level SIL-3 discussed in Section~\ref{sec:THR} below.

Therefore, we suggest a redundant channel design according to the 2oo2 pattern\footnote{In the literature, the term 
``N-out-of-M'' is used with different meanings. In this paper, NooM means that N consistent results produced by $M\ge N$ channels are needed to be accepted by the voter. Otherwise, the system falls to the safe side.} (see Fig.~\ref{fig:twochan}), where two stochastically independent sensor/perceptor implementations provide data to a voter, and the voter decides ``to the safe side'': for obstacle detection, for example, the voter would decide `obstacle present' as soon as   one channel indicates this. For distance estimates delivered by both channels in an `obstacle present' situation without disagreement, the voter selects  the shorter distance. Moreover, the voter signals  a perceptor error   to the kernel, if channels disagree over a longer time period.
To obtain stochastic independence between channels, we advocate that one channel should be implemented by conventional image processing methods, without the use of AI. Alternatively, two differently trained NNs can be used. In any case, the stochastic independence, needs to be verified by a statistical test, as described in Section~\ref{sec:independence} below.

In the remainder of this paper, the channel using conventional image evaluation techniques is denote by Channel-$\mathsf{c}$, and the channel using a NN-based perceptor by Channel-$\mathsf{n}$, as indicated in Fig.~\ref{fig:twochan}.

\section{Risk Assessment Objective and Tolerable Hazard Rate}\label{sec:THR}

The top-level hazard to be analysed  for OD is 
\begin{description}
\item[$\mathbf{H}_\mathbf{OD}$] OD signals `NO OBSTACLE' to the kernel though an obstacle is present.
\end{description}
We call this situation specified by $\mathbf{H}_\mathbf{OD}$ a \emph{false negative} produced by OD.
The objective of the risk modelling approach and the associated model evaluation by stochastic model checking is to answer the following risk analysis question.
\begin{quote}
\sl Taking into account the OD design described above: is the resulting hazard rate of $\mathbf{H}_\mathbf{OD}$ less than the tolerable hazard rate for collisions between trains and obstacles?
\end{quote}

The \emph{tolerable hazard rate (THR)} for the obstacle detection (OD) module in a freight train 
to produce a false negative (i.e.~fail to the unsafe side)
is   
\begin{equation}\label{eq:throd}
   \mathbf{THR}_\mathbf{OD} = 10^{-7}/h \quad\text{\sl -- the tolerable hazard rate for obstacle detection}
\end{equation} 
according to the discussion by Rangra et al.~\cite{rangra_analyse_2018}. This is the THR associated with   SIL-3, and it is justified by the fact that a collision between a freight train and an obstacle does not endanger as many humans, as would be the case for a passenger train. This assessment has been confirmed by the research project ATO-RISK~\cite{braband_risk_2023}, where a more detailed investigation of an adequate SIL classification for OD has been made. The project also advocates SIL-3 as the strictest safety integrity level, but additionally elaborates technical and operational boundary conditions where an even weaker SIL might be acceptable. These THR-related investigations have not yet been introduced into the current EN~5012x standards~\cite{CENELEC50126,CENELEC50126-2,CENELEC50128,CENELEC50129}, since the latter  do not consider automation grades GoA~3 or GoA~4 yet. Also, the new standard ANSI/UL~4600\xspace does not provide quantitative SIL-related requirements. It can be expected from these analyses~\cite{rangra_analyse_2018,braband_risk_2023}, however, that the ``official'' THRs, when published in new revisions of the   railway  standards, will not be stricter   
than  SIL-3 with $\mathbf{THR}_\text{OD}$ as specified in Equation~\eqref{eq:throd}. 

\section{Probabilistic Risk Assessment Approach}\label{sec:riskmodellingapproach}

The objective of the risk assessment approach described in this section is to determine a trustworthy
\emph{hazard rate} ($\mathbf{HR}_\mathbf{OD}$) for the OD module and discuss the boundary conditions ensuring that the hazard rate is less or equal to the tolerable hazard rate:
$\mathbf{HR}_\mathbf{OD} \le  \mathbf{THR}_\mathbf{OD}$.

\subsection{Strategy Overview}

\begin{figure}[t]
  \centering
  \includegraphics[width=\textwidth]{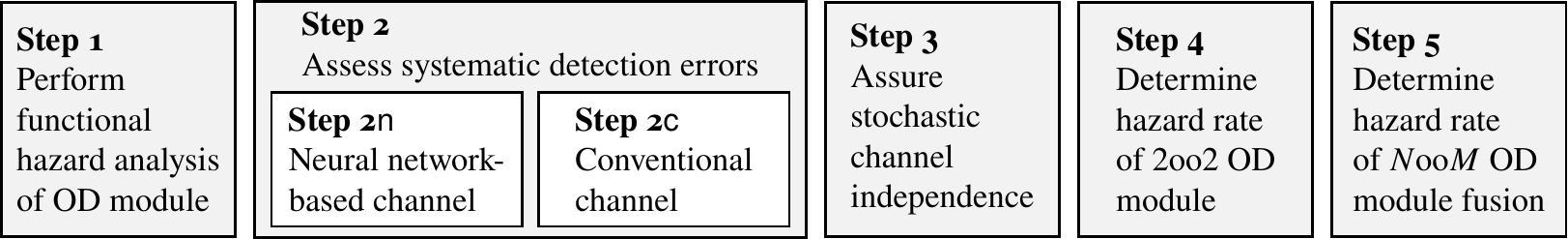}
  \caption{Overview of the probabilistic risk assessment and assurance
    approach}
  \label{fig:overview}
\end{figure}
 
The assurance strategy for the OD function comprises the following steps~(Fig.~\ref{fig:overview}).
(1)~An initial functional hazard analysis is performed for the 2oo2 OD module by means of
a fault tree analysis. This fault tree serves to check the completeness of the following bottom-up steps for risk assessment for one 2oo2 module.
(2)~The NN-based OD Channel-$\mathsf{n}$ (see Fig.~\ref{fig:twochan}) is verified by means of   statistical tests to estimate 
the  residual probability ${p_\text{\Lightning}}^\mathsf{n}$ for systematic misclassifications that may be produced by  this channel (Step~2$\mathsf{n}$ below). For OD Channel-$\mathsf{c}$ based on conventional image processing, a similar, but simpler procedure can be applied; this is   described  in Step~2$\mathsf{c}$. (3)~The stochastic independence between the two channels is demonstrated by means of another statistical test.
(4)~A continuous-time  Markov model is created for the 2oo2 OD module, and a probabilistic risk analysis is performed   by means of parametric stochastic model checking, taking the 2oo2 design into account. From this Markov model, the failure rate of the 2oo2 OD module is determined by means of stochastic model checking.
(5) With three   stochastically independent OD modules and another voter,  
a sensor/perceptor fusion can be achieved, resulting in a failure rate that conforms to the tolerable hazard rate $\mathbf{THR}_\mathbf{OD}$.
These steps are now described in detail. 

\subsection{Step~1.  Functional Hazard Analysis for OD Module}

\begin{figure}[t]
\begin{center}
\includegraphics[width=\textwidth,angle=0,origin=c]{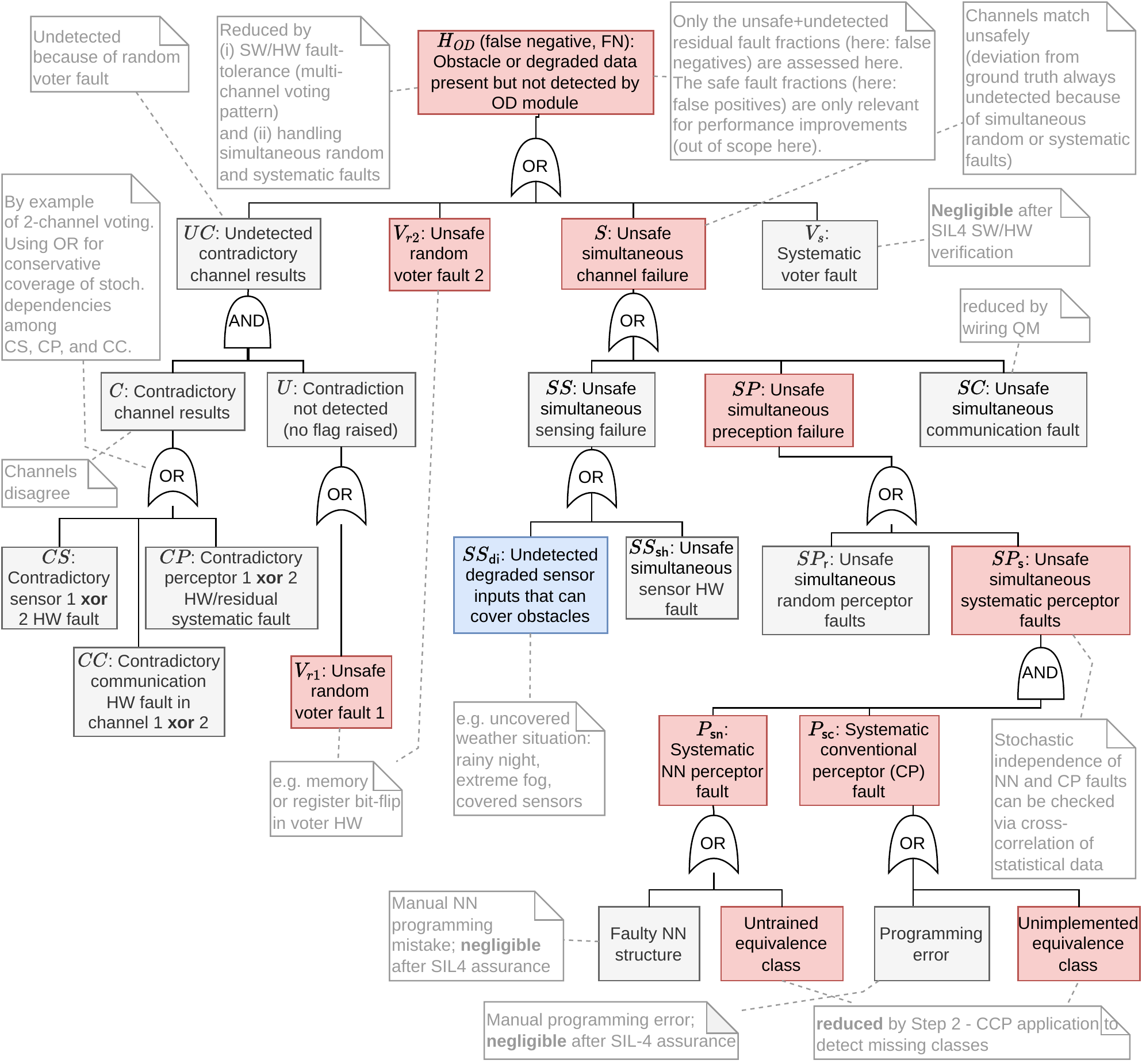}
\end{center}
\caption{Fault tree of the 2oo2-OD module for the top-level event $\mathbf{H}_\mathbf{OD}$}
\label{fig:fta}
\end{figure}

The fault tree (FT) in Fig.~\ref{fig:fta} serves as the basis for constructing the failure-related aspects and the associated mitigations in the model of the 2oo2 OD module.  We explain the most important aspects 
of the FT here.  The remaining elements of Fig.~\ref{fig:fta} should be clear from the context and the comments displayed in the figure.
The top-level hazard event $\mathbf{H}_\mathbf{OD}$ is the occurrence of a false negative (OD signals {\sl `no obstacle'} to the kernel, though an obstacle is present).

In all sub-components of the OD module (voters, sensors, perceptors, communication links, power supplies), we can assume that no systematic HW, SW or firmware failures are still present, because we require that the software is developed according to SIL-4. Therefore, the   remaining failure possibilities to consider are
(a) transient HW faults, (b) terminal HW failures, (c) systematic residual perceptor failures to detect obstacles.

The left-hand side of the FT consider cases where the two channels deliver contradictory results, but the voter fails to handle the contradiction appropriately, due to a transient fault. Undetected sensor faults (transient or terminal) in one channel can arise from HW faults or environmental conditions       (fog, snow, sandstorms).  Undetected perceptor faults can arise from HW faults or residual failures to detect certain types of obstacles.

A simultaneous channel fault (middle box on level~2) leading to $\mathbf{H}_\mathbf{OD}$ could be caused by simultaneous sensor failures  or by simultaneous perceptor faults. The former hazard is mitigated by the sensor capabilities to detect its own degradations, the stochastic independence of HW failures (due to the redundant HW design), and by the stochastic independence of the redundant perceptors, as described in Step~3 below. The latter hazard is mitigated by reducing the probability for {\it systematic} perceptor faults through the tests performed in Step~2 and by the stochastic independence of both perceptors demonstrated in Step~3, reducing the probability of a simultaneous {\it random} false negative.

\subsection{Step~2$\mathsf{n}$.  Testing  for Systematic   Classification Errors:  Channel-$\mathsf{n}$}\label{sec:classerror}

\noindent
{\bf Equivalence Classes and Their Identification -- Channel~$\mathsf{n}$}
In the real operational environment, an infinite variety of concrete obstacles could occur. Therefore, it is desirable
to partition their (finite, but still very large number of) pixel image representations into 
\emph{input equivalence classes}.
For convolutional neural networks (CNN) typically  used for image classification, it was assumed until recently that such classes could not be determined by exact calculation or at least by numerical approximation. This has changed during the last years~\cite{cheng_quantitative_2018,DBLP:journals/nn/BenfenatiM23,DBLP:journals/nn/BenfenatiM23a}, and we follow the approach of Benfenati and Marta~\cite{DBLP:journals/nn/BenfenatiM23,DBLP:journals/nn/BenfenatiM23a} for this purpose: the authors explain how to approximate the classification function of a deep NN by means of differentiable mappings $\Lambda_i$ between differentiable manifolds~$M_i$:
$$
    M_0 \stackrel{\Lambda_1}{\longrightarrow} M_1 \stackrel{\Lambda_2}{\longrightarrow} M_2\dots M_{n-1} \stackrel{\Lambda_n}{\longrightarrow} M_n
$$
Manifold $M_0$ represents the set of possible input images, and $M_1,\dots,M_{n-1}$ the intermediate hidden layers of the CNN.
For our purposes, $M_n$ is a one-dimensional output 
manifold that can be mapped to $[0,1]$, such that all $z\in[0,1)$ represent classification result {\sl ``no obstacle''}, while the $z=1$ represents {\sl ``obstacle present''}. 
The maps $\Lambda_1,\dots,\Lambda_{n-1}$ represent the inter-layer mappings of the CNN. Some of these are smooth (e.g.~the filter applications), others can be approximated by smooth alternatives.
The map $\Lambda_n : M_{n-1} \longrightarrow M_n$ is a smooth approximation\footnote{For example, by means of the Gaussian-error-linear unit (GELU).} of the ReLU (rectified linear unit) activation function, typically used in a CNN. 

Using the Euclidean metric $g_n$ on $M_n$, 
repetitive pullbacks of $g_n$ through $\Lambda_n,\dots,\Lambda_1$ introduce a {\it degenerate} Riemannian metric $g_0$ on $M_0$: using $\Lambda$ to denote the composite map $\Lambda_n\circ \dots\circ\Lambda_1 : M_0 \longrightarrow M_n$, 
the distance from $p$ to $p'$ in $M_0$ is simply $|\Lambda(p) - \Lambda(p')|$.

Given an image $p\in M_0$ that is classified by the CNN as {\sl ``obstacle''}, so that  $\Lambda(p) =  1$, all images $p'$ that can be reached from $p$ on a {\it null curve}, that is, a piecewise smooth curve of length null in the degenerate metric $g_0$, are also classified by the CNN as  obstacles.\footnote{The length of differentiable curve $\gamma$ in $M_0$ is obtained by integrating over the length of its tangent vectors is some curve parametrisation, say, $\gamma(t), t\in[0,1]$~\cite{diffgeomCarmo}. The length of a tangent vector $v = \stackrel{\cdot}{\gamma}(t)$ is obtained by calculating $\sqrt{g_0(v,v)}$: the metric $g_0$ on $M_0$ induces a bilinear form (also denoted by $g_0$ on the tangent space at $\gamma(t)$. For degenerate metrics $g_0$, non-zero tangent vectors can have zero length, since $g_0(v,v) = 0$.} The 
\emph{obstacle image space}   ${\cal O} = \Lambda^{-1} (\{ 1\}) \subseteq M_0$ of all images classified by the CNN as
obstacles, however, 
is not null-connected: for some image points $p, p''$ that are both classified as obstacles,    every piecewise smooth curve connecting $p$ and $p''$ traverses one or more regions of points mapped to {\sl ``no obstacle''}.
Each   sub-manifold  of ${\cal O}$ consisting of pairwise null-connectible image points represents an equivalence class of the CNN.

\medskip
\noindent
{\bf Statistical Test Based on Coupon Collector's Problem}
Consider   the $\ell$ equivalence classes $c_1,\dots,c_\ell \subseteq {\cal O}$ representing null-connected image sets to be classified as obstacles by   the trained NN implementing the perceptor of  Channel~$\mathsf{n}$. 
In an ideal perceptor, every real-world obstacle would produce an image  $p\in M_0$ fitting into some class $c_i$, that is, 
$p\in c_i$.
We are interested in an estimate for the residual error probability ${p_\text{\Lightning}}^\mathsf{n}$ for the existence of a further subset of ``undetetected'' images  $u_{\ell+1}\subseteq M_0 \setminus {\cal O}$ representing obstacles in the real word, but  classified as {\sl ``no obstacle''} by the NN, because they are not contained in  $\bigcup_{i=1}^\ell c_i = {\cal O}$.

Recall the naive statistical approach to estimate   ${p_\text{\Lightning}}^\mathsf{n}$: 
one could apply the Monte Carlo method by taking $n$ independent 
image samples $\{ p_1,\dots,p_n\}$ representing obstacles and determine
$
\hat{P}_{\text{\Lightning},n} = \frac{n_\text{\Lightning}}{n}$,
where $n_\text{\Lightning}$ denotes the number of false negatives obtained by   the NN on the sample  
$\{ p_1,\dots,p_n\}$. Then $\hat{P}_{\text{\Lightning},n}$ converges to ${p_\text{\Lightning}}$ for $n \rightarrow \infty$ with probability~1. This approach is unsuitable for our purposes, since the sample size $n$ must be very large for trustworthy estimation of small residual error probabilities ${p_\text{\Lightning}}^\mathsf{n}$.\footnote{Weijing Shi et al.~\cite{DBLP:conf/iccad/ShiALYAT16} state that  
a misclassification probability of ${p_\text{\Lightning}} \approx 10^{-12}$ would require a sample size of  $n \approx 10^{13}$.}

As a more promising alternative approach, we therefore suggest to obtain   an    estimate for ${p_\text{\Lightning}}^\mathsf{n}$ by means of statistical tests based on a generalised variant of the  \emph{Coupon Collector's Problem (CCP)}~\cite{FLAJOLET1992207}. This CCP variant considers $\ell$ different types of coupons in an urn, such that drawing a coupon of type $i \in \{ 1,\dots,\ell\}$ from the urn with replacement has probability $p_i$.   The CCP considers the random variable $X$ denoting the number of draws necessary to obtain a coupon of each type at least once.
The expected value of $X$ is calculated by~\cite{FLAJOLET1992207}
\begin{equation}\label{eq:expected}
E(X) = \int_0^\infty \big( 1 - \prod_{i=1}^\ell (1 - e^{-p_i x})  \big)\mathrm{d}x\;.
\end{equation}

For the application of the CCP in the context of this article, we assume  the availability of a large database $D$ of `obstacle-on-track' sample images representing the urn in the CCP. We assume that there exists a random selection mechanism for $D$, such that the selected samples are stochastically independent from each other, concerning their ontology classification. The obstacle images from $D$ take the role of the CCP-samples to be drawn from the urn. If the image sample fits into equivalence class $c_i, \ i\in \{1,\dots,\ell\}$, this corresponds to the CCP coupon being of type $i$.

For a verification run $V_k$, we draw sample images from $D$ until every known equivalence class $c_1,\dots,c_\ell$ has been covered by at least one sample. If a run $V_k$ fails because a substantial number of samples did not fit into any class, the training of the neural network is repeated with an extended set of samples, and a new collection of equivalence classes $c_1',\dots c_{\ell'}'$ is determined, as described  above. Then the verification runs are repeated.  

While $E(X)$ gives us an idea of   the number of samples needed to cover every equivalence class $c_i$ at least once,  we need a (higher) value of samples $\overline S$ required to discover all members {\it with sufficiently high probability}. Hence we are looking for an   $\overline S\in \mathbb{N}$ such that the
probability    $\tau = \mathsf{P}(X < \overline S)$ is close to~1. Adapting the estimation approach suggested  by Hauer  et al.~\cite{DBLP:conf/itsc/HauerSHP19} to our problem,  the probability $\mathsf{P}(X < S)$ for any   $S\in \mathbb{N}$ can be calculated by using a large number of verification runs $V_k,\ k = 1,\dots,m$  and count the occurrences   $occ(i),\ i \in \mathbb{N}$ of verification runs in 
$\{V_1,\dots,V_m\}$, where all equivalence classes
$c_1,\dots,c_\ell$ have been covered after $i$ samples. Note that $occ(i) = 0$ for $i < \ell$, because we need at least $\ell$ images to cover that many classes. 
Then   $\mathsf{P}(X < S)$ can be estimated by
\begin{equation}\label{eq:Sestimate}
   \mathsf{P}(X < S) = \frac{1}{m}\sum_{i=1}^S occ(i)\;,
\end{equation}
so we select $\overline S$ as the smallest $S$ such that the value of $\mathsf{P}(X < S)$ calculated by Equation~\eqref{eq:Sestimate} is greater or equal $\tau$. The number $m$  of verification sets $V_k$ to be used in Equation~\eqref{eq:Sestimate} determines the confidence that we can have in the estimate for $\mathsf{P}(X < \overline S)$; minimal values  for $m$ achieving a given confidence level can be calculated as described by Hauer  et al.~\cite{DBLP:conf/itsc/HauerSHP19}.

Assume that the NN implementing the perceptor of Channel~$\mathsf{n}$ has $\ell$ equivalence classes, as described  above. Assume further that the verification runs $V_k,\ k=1,\dots,m$ have been performed successfully and resulted in probability estimates $p_1,\dots,p_\ell$ for an obstacle to fall into class $c_1,\dots,c_\ell$.
Now we test the hypothesis that the successful verification runs have overlooked an obstacle type that does not fit into any of the identified classes $c_1,\dots,c_\ell$, but is associated with a subset $u\subseteq M_0\setminus {\cal O}$ of obstacle images leading to false negatives.  Assume further that the occurrence probability 
for such an obstacle is $p_u$. Then we have to re-scale the probability estimates $p_1,\dots,p_\ell$   
to $p_i' = (1-p_u)p_i$, so that
$
\big(\sum_{i=1}^\ell p_i'\big) + p_u = 1.
$
According to Equation~\eqref{eq:expected},
the expected number of   samples needed for covering the $(\ell+1)$ classes is
$$
E(X) = \int_0^\infty \Big( 1 - \big(\prod_{i=1}^\ell (1 - e^{-p_i x})\big)\cdot (1 - e^{-p_u x})  \Big)\mathrm{d}x\;.
$$
Now we apply Equation~\eqref{eq:Sestimate} for this extended hypothetical     partition
$\{c_1,\dots,c_\ell, u\}$,   to estimate the number $m_\text{new}$ of verification runs $V_k$ to be performed in order to get at least one sample for each partition element, {\it including} $u$, again with a high confidence level and the same probability $\mathsf{P}(X_\text{new} < \overline S_\text{new}) = \tau$. Then the verification runs are extended to 
$V_1,\dots,V_{m_\text{new}}$. If this extended suite of verification runs does {\it not} reveal the existence of 
such a partition element $u$,   we can conclude with the given confidence level that the original set of classes $c_1,\dots,c_\ell$ implemented by the NN is complete.

\subsection{Step~2$\mathsf{c}$.  Testing  for Systematic   Classification Errors:   Channel-$\mathsf{c}$}\label{sec:classerrorc}

\medskip
\noindent
{\bf Equivalence Classes and Their Identification}
For the perceptor of Channel~c, an input equivalence class consists of a set of images covering   the same path in the perceptor software control flow graph, so that 
they all end up with the same classification result. 

\medskip
\noindent
{\bf Statistical Tests}
The statistical tests regarding the probability ${p_\text{\Lightning}}^\mathsf{c}$ of systematic residual classification errors in Channel-$\mathsf{c}$ can be
performed in analogy to Step~2$\mathsf{n}$, but now the equivalence classes are identified by software control flow paths instead of null-connected sub-manifolds of the obstacle image space ${\cal O}$. 

\subsection{Step~3. Stochastic Independence Between the two Channels}\label{sec:independence}

On hardware-level, stochastic independence between the two OD channels  is justified by  redundancy and segregation arguments: the channels use different cameras, and the perceptors  
are deployed on different processor boards with separate power supplies and wiring, both for electrical current and communication links between sensors, perceptors, and voter. There are no communication or synchronisation links between the channels.

The remaining common cause failure of the two channels that cannot be avoided is given by adverse weather conditions (like fog, sand storms, or  snow) corrupting the camera images. This can be detected by the sensors themselves by identifying consecutive images as identical without discernible shapes (fog) or as white noise (sand storm, snow). We can expect that at least one of the two channels detects this condition and raises a fault that will cause the voter to signal `OD failure' to the kernel. This will lead to an emergency stop of the train.  
Consequently, we are only interested in stochastic independence of the two perceptors {\it in absence} of this detectable common cause failure.

As discussed for the fault tree model of Step~1, the only remaining potential cause for   stochastic dependence would be that the two perceptors evaluate images ``in a similar way''.
To demonstrate the absence of such a dependency, we apply the method of Sun et al.~\cite{DBLP:conf/eccv/SunCHK20} for explaining the {\it reasons} for classification results: the method provides an algorithm for identifying a subset of image pixels that were essential for obtaining the classification result. For the demonstration of stochastic independence, we
define  two  bit matrix-valued random variables $R_i,\ i = \mathsf{c},\mathsf{n}$. 
Variable $R_i$ encodes these explanations obtained by Channel~$\mathsf{c}$ and Channel~$\mathsf{n}$, respectively, as  a pixel matrix, where only the essential pixels are represented by non-zero values. 

While performing the verification runs $V_k$ of Step~2c and Step~2n, the sequence of matrix values $R_\mathsf{c}, R_\mathsf{n}$ obtained from the images of $V_k$ are determined (both channels need to run the same verifications $V_k$ in the same order, so that the same sequence of images is used over all runs $V_k$). Then the stochastic independence between $R_\mathsf{c}$ and $R_\mathsf{n}$ can be tested by means of the
widely-used $\chi^2$-test. If this test indicates a stochastic {\it dependence} between perceptors~$\mathsf{c}$ and $\mathsf{n}$, then the NN has to be retrained with a different data set, or another structure of the NN (for example, another layering) needs to be chosen.  

The main result obtained from the stochastic independence of $R_\mathsf{c}$ and $R_\mathsf{n}$ is that false negative misclassifications occur at the two channels in a stochastically independent way. More formally, let $X_i,\ i = \mathsf{c},\mathsf{n}$ be two Boolean random variables with interpretation 
``$X_i = \text{true}$ if and only if a  false negative misclassification occurs in the perceptor of  Channel~$i$''.
Then, with $a,b \in \{ \text{true}, \text{false} \}$, stochastic independence allows us to calculate
$$
  \mathsf{P}(X_\mathsf{c}= a \land X_\mathsf{n}=b) =
  \mathsf{P}(X_\mathsf{c}=a) \cdot \mathsf{P}(X_\mathsf{n}=b)\;.
$$
In particular, the probability of a simultaneous misclassification in both channels (case $a = \text{true}~\wedge~b = \text{true}$) can be calculated as the product of the misclassification probabilities of each channel.

\subsection{Step~4. Determining $\mathbf{HR}_\mathbf{OD}$ for the 2oo2-OD
  Module}\label{sec:tle-calc}

We now quantify the probability of an $\mathbf{H}_\mathbf{OD}$ event for a single
module demand. 
Recall that $\mathbf{H}_\mathbf{OD}$ means an obstacle is
present within OD range or the OD module is provided with degraded
data, but neither is detected by the module (i.e., $r=\mathrm{no}$) and
the module's voter component misses to raise an error flag~(i.e.,
$f=\mathrm{false}$) that could be considered by the automatic train protection.

\begin{wrapfigure}[20]{r}{.55\linewidth}
  \vspace{-2em}
  \includegraphics[width=\linewidth]{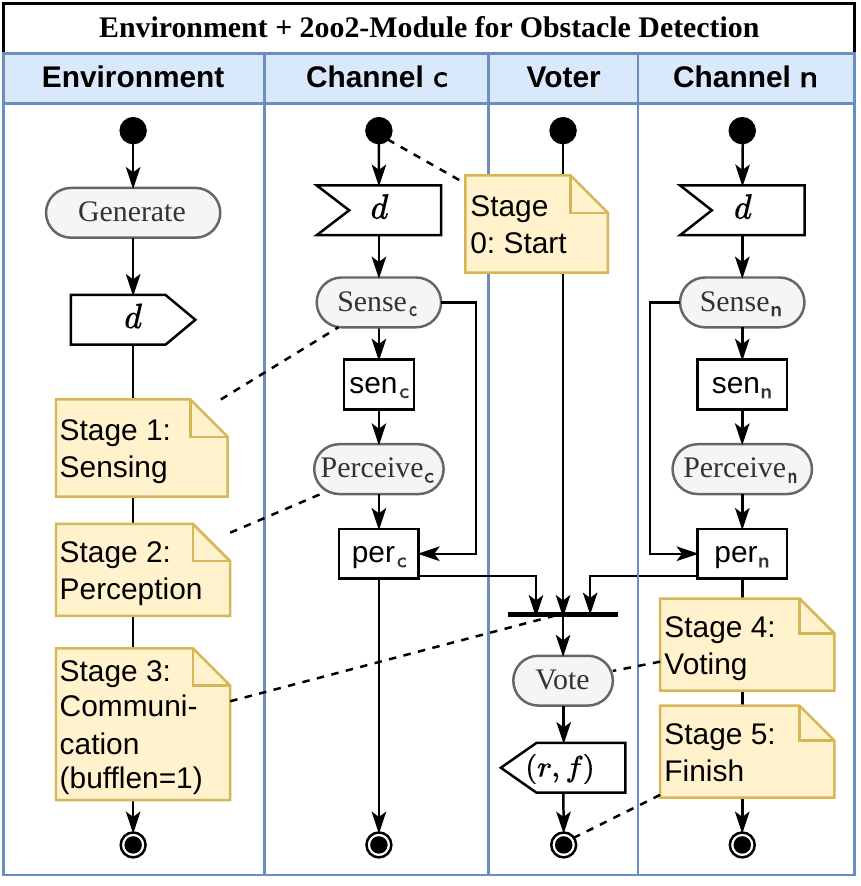}
  \caption{SysML activity chart describing the data
    processing in the OD module}
  \label{fig:atrain-ftpattern}
\end{wrapfigure}
We model the two channels~($i\in\{\mathsf{c,n}\}$) and the voter of
the OD module~(Fig.~\ref{fig:twochan}) using a SysML activity chart
with a Petri net interpretation as shown in
Fig.~\ref{fig:atrain-ftpattern}.  For each processing cycle (i.e.,
when new tokens are placed at the beginning of each line), both
channels perform a \emph{sense} and a \emph{perceive} action with the
data $d\in D$ flowing (i.e., carried with the tokens) from the
environment into both channels and from the top to the bottom.  For
illustration, we use $D=0..2$, with $d=0$ for ``obstacle absent'',
$d=1$ for ``obstacle present'', and $d=2$ for ``degraded inputs''
(e.g., dense fog, covered sensors).  The \emph{environment} part
enables a conditional risk assessment of the OD module based on the
stochastic \emph{generation} of inputs from~$D$.  In our assessment,
the environment only generates $d\in\{1,2\}$.

We use a continuous{\Hyphdash}time Markov chain~(CTMC) as stochastic
model for the OD module.  Given variables $V$, a CTMC is a tuple
$\mathcal{M} = (S,s_0,\mathbf{R},L)$ with state space
$S\in 2^{V\to\mathbb{N}}$, initial state $s_0\in S$, transition rate matrix
$\mathbf{R}\colon S\times S\to\mathbb{R}_{\geq 0}$, and labelling
function $L\colon S\to 2^{\mathit{AP}}$ for a set $\mathit{AP}$ of
atomic propositions.
Properties to be checked of $\mathcal{M}$ can be specified in
continuous stochastic logic~(CSL).  For example, the expression
$\mathcal{M},s \models \mathsf{P}_{>p}[\mathsf{F} \phi]$ is satisfied
if and only if the CSL formula $\mathsf{P}_{>p}[\mathsf{F} \phi]$ is
true in $\mathcal{M}$ and $s\in S$, that is, if the probability
($\mathsf{P}$) of eventually ($\mathsf{F}$) reaching some state $s'$
satisfying $\phi$ from $s$ in $\mathcal{M}$ is greater than~$p$.  If
$\phi$ is a propositional formula, its satisfaction in $s\in S$
($s\models\phi$) is checked using the atomic propositions from $L(s)$.
More details about CSL model checking, for example, with the
\textsc{Prism} model checker, can be obtained
from~\cite{Kwiatkowska2011-PRISM4Verification}.

\begin{figure}[htb]

\includegraphics[width=\linewidth]{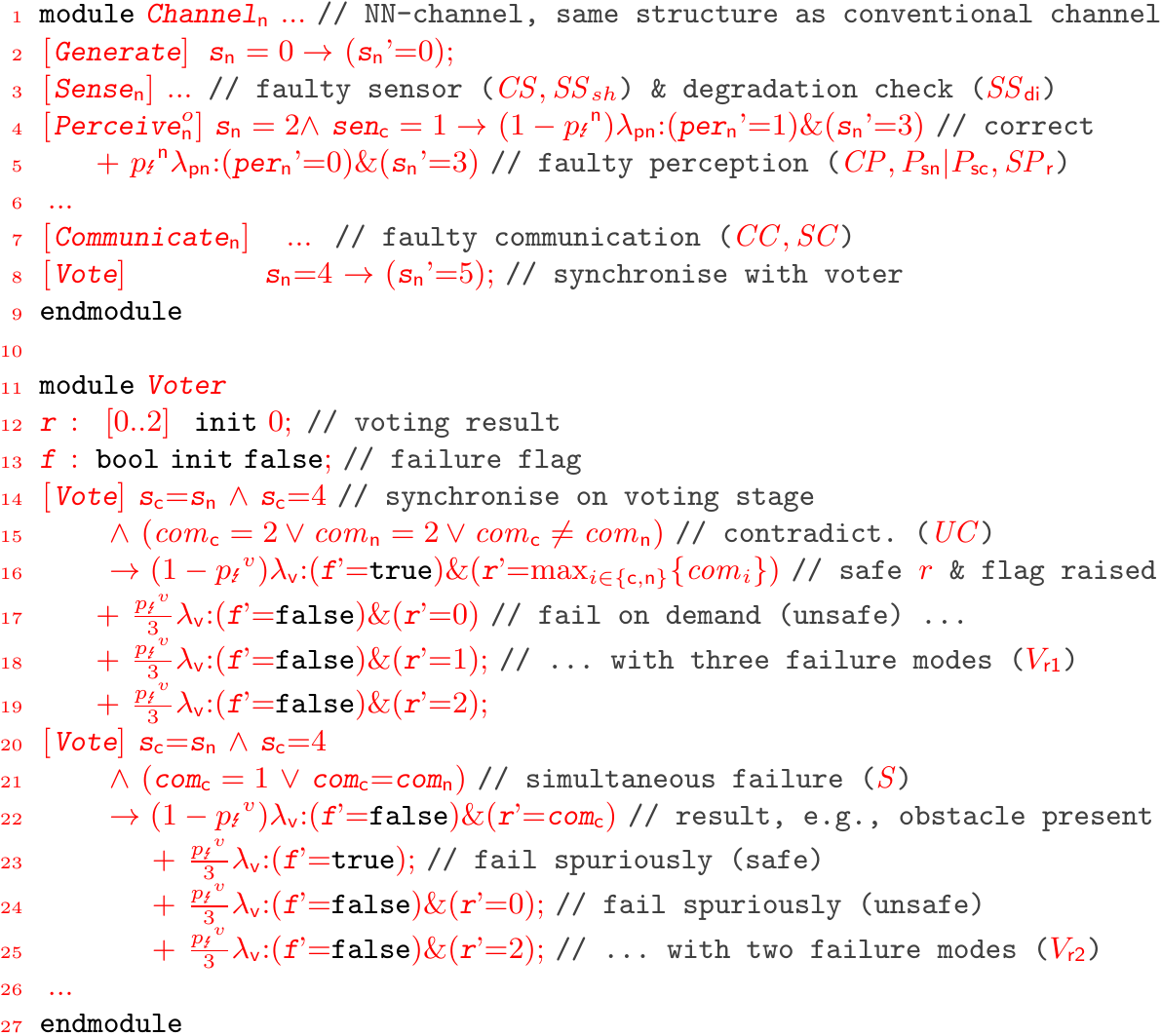}
\caption{Probabilistic program fragment showing parts of the NN channel and the voter. The influence on some of the FT events from Fig.~\ref{fig:fta} is indicated.}
\label{lst:ctmc-voter}
\end{figure}

To work with CTMCs, we translate the activity chart from
Fig.~\ref{fig:atrain-ftpattern} into a probabilistic guarded
command\footnote{Such commands are of the form
  $[a] g\to \lambda_1\colon u_{11}\&u_{12}\dots + \dots + \lambda_{n}\colon
  u_{nm}\dots$ with an action $a$, a guard $g$, and probabilistic
  multiple-assignments $u_{ij}$ applied with rate $\lambda_i$.}
program~(Listing~\ref{lst:ctmc-voter}).  From this program, a
probabilistic model checker can derive a CTMC $\mathcal{M}$ that
formalises the semantics of the activity chart, allowing the
processing in the two channels to be non-deterministically
concurrent,\footnote{
  Each of the timed synchronised interleavings of the four sequential
  components in Fig.~\ref{fig:atrain-ftpattern} carries information
  about the \emph{expected time of occurrence} of events and, thus,
  the accumulated expected duration of a particular interleaving.
  This allows one to derive timed termination probabilities and rates
  of the processing cycle.}  finally synchronising on the \emph{vote}
action.  This type of concurrency enables us to make assumptions about
the processing speed in the two channels independently and flexibly.

The Listing~\ref{lst:ctmc-voter} shows fragments of the program
describing one channel, its processing stages, and the voter
component.  The state space $S$ of the associated $\mathcal{M}$ is
defined via a stage counter ($s_i\in 0..5$), data flow variables
($\mathit{sen}_i$, $\mathit{per}_i$, $\mathit{com}_i : D$) for each
channel, variables for the input data $d: D$, the result $r: D$, and a
Boolean failure flag $f$.  We use the initial state $s_0(v)=0$ for
$v\neq f$ and $s_0(f)=\mathrm{false}$.
The transition rate matrix $\mathbf{R}$ is defined indirectly via
probabilistic updates: For each update $u$~(e.g., a fault) of an
action $a$~(e.g., $\mathtt{Perceive}_{\textsf{n}}^o$), we provide a rate
$\lambda_{a,u} = p_u \cdot \lambda_a$, where $p_u$ is the probability
of seeing update $u(s)$ if an action $a$ is performed in state $s$ and
$\lambda_a$ is the average speed, frequency, or rate at which action
$a$ in $s$ is completed.  We can either provide a compound rate
$\lambda_{a,u}$ or separate parameters $p_u$ and $\lambda_a$.  For
example, for the $\mathtt{Perceive}_{\textsf{n}}^o$ action (i.e.,
NN-based perception, given the sensor forwards a picture with an
obstacle, line~4), we consider a single failure mode (line~5) with probability
${p_\text{\Lightning}}^\mathsf{n}$ (estimated in Sect.~\ref{sec:classerror}) multiplied
with a perception speed estimate $\lambda_{\mathsf{pn}}$.

As described in Sect.~\ref{sec:background}, the output at the end of
each processing cycle is a tuple~$(r,f)$ with the voting result~$r$
and the status of the failure flag~$f$.
Under normal operation, $r$ contains either the concurring result of
both channels or an error to the \emph{safe side} (i.e.,
$\max_{i\in\{\mathsf{c,n}\}}\{\mathit{com}_i\}$) in case of
contradictory channel results.  For example, if one channel reports an
obstacle and the other does not, the nominal voter would forward
``obstacle present'' and raise a flag.

For the model, we need to provide probability and speed estimates of
the channel- and stage-specific faults.  For example, we use
${p_\text{\Lightning}}^\mathsf{n}$ and $\lambda_{\mathsf{pn}}$ for the probability of an
NN-perceptor fault $\mathit{SP}_{\mathsf{n}}$ and the
speed\footnote{Speed estimates can be set to 1 for a CTMC where
  estimates are unavailable and relative speed and performance does
  not play a role in the risk assessment.} of the associated
fault-prone action $\mathtt{Perceive}_{\mathsf{n}}^o$.  Analogously, we
provide ${p_\text{\Lightning}}^\mathsf{c}$ and $\lambda_{\mathsf{pc}}$ for the conventional
perceptor, ${p_\text{\Lightning}}^v$ and $\lambda_{\mathsf{v}}$ for $V_{\mathsf{r}}$, and,
similarly, for the other events defined in the fault tree (e.g.,
$\mathit{SP}_{\mathsf{r}}$, $\mathit{SP}_{\mathsf{s}}$, $\mathit{SC}$,
$\mathit{SS}_{\mathsf{di}}$, $\mathit{SS}_{\mathsf{sh}}$;
Fig.~\ref{fig:fta}).  Based on these parameters, the CTMC allows us to
quantify time-independent probabilities of intermediate and top-level
events in the fault tree, for example, $\mathit{UC}$, $S$, and, in
particular, the probability $\mathsf{P}[\mathit{FN}]$ of the top-level
event~$\mathbf{H}_\mathbf{OD}$, that is, a false negative under the condition that
either an obstacle or degraded data is present.

To make our assessment independent of a particular ${p_\text{\Lightning}}^\mathsf{n}$ and
${p_\text{\Lightning}}^\mathsf{c}$, we perform a parametric CTMC analysis that yields a
function $\mathsf{P}[\mathit{FN}]({p_\text{\Lightning}}^\mathsf{n},{p_\text{\Lightning}}^\mathsf{c})$.
Consider the parametric CTMC
$\mathcal{M}({p_\text{\Lightning}}^\mathsf{n},{p_\text{\Lightning}}^\mathsf{c}) =
(S,s_0,\mathbf{R}({p_\text{\Lightning}}^\mathsf{n},{p_\text{\Lightning}}^\mathsf{c}),L)$ derived from
Listing~\ref{lst:ctmc-voter}.
By
$S_{\mathsf{od}}=\{s\in S\mid s_{\mathsf{c}}=s_{\mathsf{n}} \land
s_{\mathsf{c}}=1 \land (d=1\lor d=2)\}$, we select only those
\emph{intermediate states} where the OD module is provided with either
a present obstacle ($d=1$) or degraded data ($d=2$) at its sensing
stage ($s_{\mathsf{c}}=1$).  According to the fault
tree~(Fig.~\ref{fig:fta}), we select \emph{final states} with the
predicate
\begin{align*}
  \mathrm{fin} &\equiv\; \big((s_{\mathsf{c}}=s_{\mathsf{n}} \land s_{\mathsf{c}}=5 \land\neg f)
    \tag*{at final stage $s_i=5$ with muted flag ($V_{\mathrm{r}}$),}
  \\&\land\; ((com_{\mathsf{c}}\neq com_{\mathsf{n}})
  \tag*{observe contradictory results ($UC$), \textbf{or} a}
  \\&\lor
  (com_{\mathsf{c}}=com_{\mathsf{n}}\land r \neq d))
  \big)
  \tag*{simultaneous channel or voter fault ($S,V_r$).}
\end{align*}
These are all states at the final processing stage ($s_i=5$) that
correspond to either $\mathit{UC}$ or $S$ in the fault tree and,
hence, $\mathbf{H}_\mathbf{OD}$.
Then, we compute $\mathsf{P}[\mathit{FN}]({p_\text{\Lightning}}^\mathsf{n},{p_\text{\Lightning}}^\mathsf{c})$ by
quantifying ($\mathsf{P}_{=?}[\cdot]$) and accumulating
($\sum_{S_0}\cdot$) the conditional probabilities of the unbounded
reachability ($\mathsf{F}\;\mathrm{fin}$) of a final state in
$S_{\mathsf{f}}=\{s\in S\mid s\models\mathrm{fin}\}$ from some
intermediate state~$s\in S_{\mathsf{od}}$.  The corresponding formula is
\begin{align}
  \label{eq:prob-fail-on-dem}
  &\mathbf{H}_\mathbf{OD}({p_\text{\Lightning}}^\mathsf{n},{p_\text{\Lightning}}^\mathsf{c})
  = \mathsf{P}[\mathit{FN}]({p_\text{\Lightning}}^\mathsf{n},{p_\text{\Lightning}}^\mathsf{c})
  \\&= \sum_{s\in S_{\mathsf{od}}}
  \big(\underbrace{
  \mathcal{M}({p_\text{\Lightning}}^\mathsf{n},{p_\text{\Lightning}}^\mathsf{c}),s_0\models\mathsf{P}_{=?}[\mathsf{F}\,s]
  }_{\text{probability of reaching $s$ from $s_0$}}\big)
  \cdot
  \big(\underbrace{\mathcal{M}({p_\text{\Lightning}}^\mathsf{n},{p_\text{\Lightning}}^\mathsf{c}),s
  \models \mathsf{P}_{=?} [ \mathsf{F}\,\mathrm{fin}
  ]}_{\text{probability of reaching $\mathrm{fin}$ from $s$}}\big)\;.
  \notag
\end{align}
Note that the CSL quantification operator $\mathsf{P}_{=?}$ used
inside the sum operator transforms the satisfaction relation $\models$
into a real-valued function.

\begin{figure}[t]
  \subfloat[{$\mathsf{P}[\mathit{FN}]({p_\text{\Lightning}}^\mathsf{n},{p_\text{\Lightning}}^\mathsf{c})$}]{
    \label{fig:param-pfd-assess}
    \includegraphics[width=.48\linewidth]{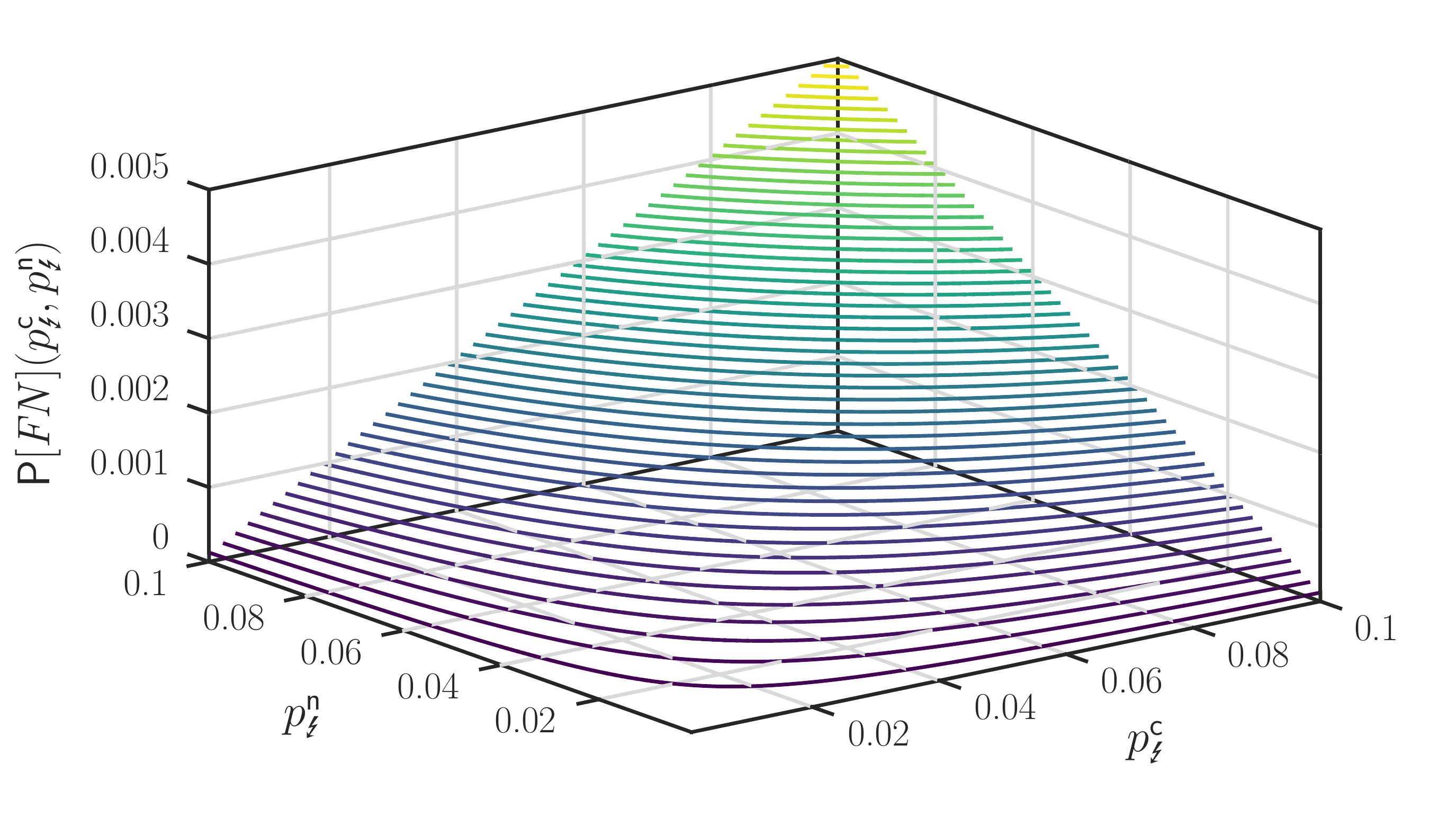}}
  \subfloat[{$\mathbf{HR}_\mathbf{OD}({p_\text{\Lightning}}^\mathsf{n},{p_\text{\Lightning}}^\mathsf{c})$}]{
    \label{fig:param-hr-assess}
    \includegraphics[width=.48\linewidth]{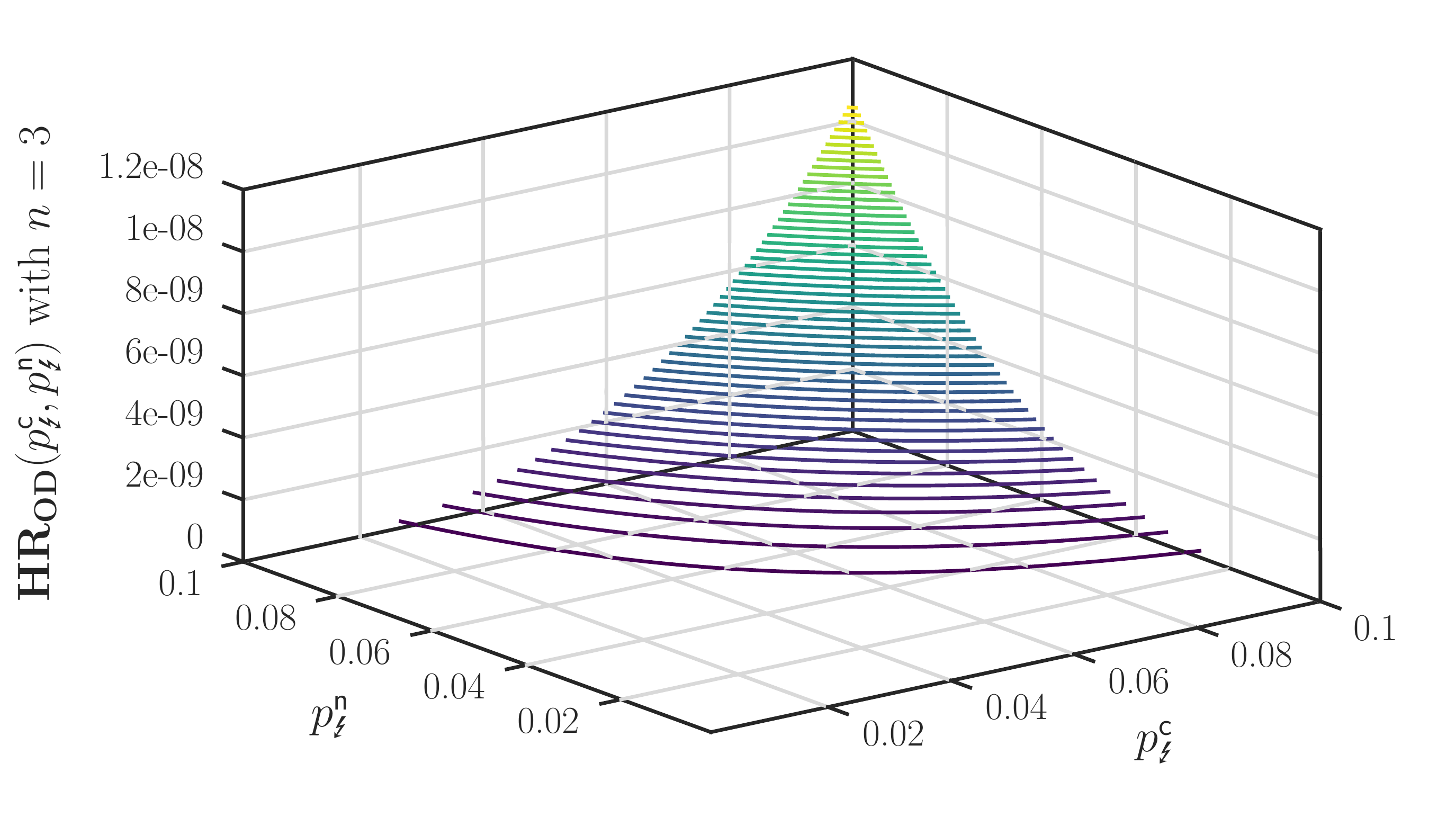}}
  \caption{The functions in (a) and (b) result from computing the symbolic solution
    of the right-hand side of Eq.~\eqref{eq:prob-fail-on-dem} using the
    parametric CTMC $\mathcal{M}({p_\text{\Lightning}}^\mathsf{n},{p_\text{\Lightning}}^\mathsf{c})$.}
  \label{fig:param-risk-ass-example}
\end{figure}

Shown in Fig.~\ref{fig:param-pfd-assess}, one OD module has a residual probability for an undetected false negative in range $\mathsf{P}[\mathit{FN}]({p_\text{\Lightning}}^\mathsf{n},{p_\text{\Lightning}}^\mathsf{c}) \in [0.0016,0.005]$, depending on the residual 
misclassification probability ${p_\text{\Lightning}}^\mathsf{n},{p_\text{\Lightning}}^\mathsf{c}\in [0.02,0.1]$. Reports on failure probabilities of image classification based on both conventional image evaluation and trained neural networks indicate that, as of today,
neither ${p_\text{\Lightning}}^\mathsf{n}$ nor ${p_\text{\Lightning}}^\mathsf{c}$ will be below 
$0.02$~\cite{ristic-durrant_review_2021,DBLP:journals/nn/AkaiHM21}. For example, assuming ${p_\text{\Lightning}}^\mathsf{n} = {p_\text{\Lightning}}^\mathsf{c} = 0.04$, one OD module alone
will have a hazard rate of approximately  $\lambda_{\mathsf{od}} \cdot 0.0016 = 6\cdot 10^{-5}$ with $\lambda_{\mathsf{od}} = 2/24h^{-1}$ denoting the frequency of obstacle occurrences or degraded data. While this does not yet conform to 
$\mathbf{THR}_\mathbf{OD}$ specified in Equation~\eqref{eq:throd} (see also the parameter-dependent hazard rates in Fig.~\ref{fig:param-pfd-assess}), it allows to apply sensor fusion to create a composite OD system respecting $\mathbf{THR}_\mathbf{OD}$.

\subsection{Step~5. Determining $\mathbf{HR}_\mathbf{OD}$ for the Fused 3oo3 OD System}\label{sec:fusion}

We create a 3oo3 sensor fusion system, using three stochastically independent (that is, differently trained and with different image recognition software) OD modules with 2-channel structure as described above: 
a 3oo3 voter raises an error leading immediately to braking the train, as soon as an {\sl ``obstacle/no obstacle''} indication is no longer given unanimously by the three OD modules. This means that single and double faults are immediately detected and result in immediate fault negation by going into a safe state.
As explained in the previous paragraph, each module has a failure rate that is smaller than $2\cdot 10^{-4}h^{-1}$.
Therefore, applying the rule~\cite[B.3.5.2, 5)]{CENELEC50129} of EN~50129, the detection of triple faults for such a system is not required.

Assuming that all three OD modules have a probability for producing a false negative that is less or equal to $\mathsf{P}[\mathit{FN}]({p_\text{\Lightning}}^\mathsf{n},{p_\text{\Lightning}}^\mathsf{c})$, the  hazard rate for a safety-critical false negative produced by this 3oo3 OD system (Fig.~\ref{fig:param-hr-assess}) is  
\begin{equation}
  \label{eq:hazard-rate}
  \mathbf{HR}_\mathbf{OD}({p_\text{\Lightning}}^\mathsf{n},{p_\text{\Lightning}}^\mathsf{c})
  = \lambda_{\mathsf{od}}
  \cdot \big(\mathsf{P}[\mathit{FN}]({p_\text{\Lightning}}^\mathsf{n},{p_\text{\Lightning}}^\mathsf{c})\big)^3 \;.
\end{equation}
With $\mathsf{P}[\mathit{FN}](0.04,0.04) = 0.0016$ as discussed above, this ensures
$$
   \mathbf{HR}_\mathbf{OD}(0.04,0.04) = \frac{2}{24} \cdot 0.0016^3 = 3.413\cdot 10^{-10} < \mathbf{THR}_\mathbf{OD} = 10^{-7}\;.
$$

\section{Conclusion}\label{sec:conc}

We have presented a   5-step approach to probabilistic risk assessment for autonomous freight trains in open environments with automated obstacle detection. This approach is based on a preceding qualitative evaluation of the assurance steps required to 
enable a certification according to the standard ANSI/UL~4600\xspace. The risk figures obtained indicate that autonomous freight trains based on the train control system architecture advocated here can achieve adequate safety with obstacle detection based on camera images alone,  provided that at least three independent 2oo2 OD modules are fused into an integrated 3oo3 OD detector. 
The costs to achieve this  can be expressed in the number of statistical tests to be performed in order to guarantee these upper risk bounds. 
Moreover, our example illustrates that, under realistic assumptions, the failure probability of an OD module corresponds to the product of the classification error probabilities; sensor and voter faults play no significant role in the overall assessment.

The statistical testing strategy described here   requires considerable effort, since several verification runs $\{ V_1,\dots,V_{m_\text{new}} \}$ are involved and have to be repeated if too many false negatives require a new training phase. To avoid the latter, it is advisable to verify first that the trained NN is \emph{free of adversarial examples}: in our case, these are images $p, p'$ that are close to each other in some metric conforming to the human understanding of image similarity (e.g.~two similar vehicles standing on the track at a level crossing), where $p$ is correctly classified as an obstacle, but $p'$ is not. A highly effective testing method for detecting adversarial examples has been suggested by Sun et al.~\cite{sun_concolic_2018}. It is based on a novel structural coverage metric for CNN, that is analogous to the MC/DC coverage in software testing. A detailed verification  cost evaluation will be considered in a future contribution.

It is important to note that the introduction of redundancy (e.g.~2oo2) to achieve fail-safe designs, as described in EN~50129~\cite[B.3.1]{CENELEC50129}, is only admissible for random HW faults according to this standard. The occurrence of residual HW design failures, SW failures, or failures due to imperfect machine learning processes is not taken into account. For HW design and SW (including the implementation of NN software) developed and verified according to SIL-4, the assumption that safety-critical residual failures exist can be also neglected for the context of this paper. The probability of a residual systematic failure in a trained NN, however, needs to be taken into account. Therefore, a certification of 
the OD module in an autonomous freight train cannot be performed on the basis of the current EN~5012x standards alone. Instead, ANSI/UL~4600\xspace needs to be used: according to this new standard for autonomous control systems, the failure model is allowed to take systematic residual failures caused by imperfect machine learning into account.

\bibliographystyle{splncs03}
\bibliography{main-arxiv}

\end{document}